# Reduction of high reset currents in unipolar resistance switching Pt/SrTiO$_x$/Pt capacitors using acceptor doping

S. B. Lee,[1] A. Kim,[1] J. S. Lee,[2] S. H. Chang,[1] H. K. Yoo,[1] T. W. Noh,[1] B. Kahng,[2] M.-J. Lee,[3] C. J. Kim,[3] and B. S. Kang[4,a]

[1]*ReCOE, Department of Physics and Astronomy, Seoul National University, Seoul 151-747, Republic of Korea*

[2]*Department of Physics and Astronomy, Seoul National University, Seoul 151-747, Republic of Korea*

[3]*Samsung Advanced Institute of Technology, Yongin, Gyeonggi-do 446-712, Republic of Korea*

[4]*Department of Applied Physics, Hanyang University, Ansan, Gyeonggi-do 426-791, Republic of Korea*

[a]Electronic mail: bosookang@hanyang.ac.kr

The high reset current, $I_R$, in unipolar resistance switching is an important issue which should be resolved for practical applications in nonvolatile memories. We showed that, during the forming and set processes, the compliance current, $I_{comp}$, can work as a crucial parameter to reduce $I_R$. Doping with Co or Mn can significantly reduce the leakage current in capacitors made using $SrTiO_x$ film, opening a larger operation window for $I_{comp}$. By decreasing $I_{comp}$ with acceptor doping, we could reduce $I_R$ in $SrTiO_x$ films by a factor of approximately 20. Our work suggests that the decrease of $I_{comp}$ by carrier doping could be a viable alternative for reducing $I_R$ in unipolar resistance switching.

Resistance switching (RS) is a physical phenomenon where reversible changes between two bistable resistance states can be achieved by the application of an external bias.[1-5] In unipolar RS, the switching conditions depend on the magnitude of the bias voltage irrespective of the polarity. Recently, unipolar RS has attracted renewed interest due to potential applications in nonvolatile memory devices, known as resistance random access memory (RRAM).[3-10] However, several scientific and technical issues remain to be resolved prior to the commercial realization of the technology.

One such issue is the reduction of the reset current, $I_R$.[6-8] During the reset process, i.e., a change from the low resistance state (LRS) to the high resistance state (HRS), there will be a large current in the RRAM device. By reducing $I_R$, however, the power consumption of the device will be reduced. In addition, to fabricate a high-density three-dimensional RRAM array structure, diodes and/or transistors are required as switch elements.[11] If $I_R$ can be reduced, the requirements of the switches can become less demanding, enhancing the feasibility of practical three-dimensional RRAM devices. There have been a number of efforts to reduce $I_R$, for example using a nanowire as the switching material,[6] by reducing the dimensions of the electrode,[7] or reducing the parasitic capacitance in capacitor-type RRAM devices.[8] Here, we show that acceptor doping of $SrTiO_x$ (STO) can significantly reduce $I_R$ in Pt/STO/Pt capacitors.

We used polycrystalline STO thin films as the unipolar RS material. We prepared undoped STO, Co-doped STO (Co:STO), and Mn-doped STO (Mn:STO) thin films on Pt-coated Si substrates. We deposited polycrystalline films 60 nm-thick using pulsed laser deposition. The targets were stoichiometric STO, 0.5 % wt. Co:STO, and 0.5 % wt. Mn:STO, respectively. The fabrication conditions were as follows: a substrate temperature of 600 °C, an oxygen pressure of 100 mTorr, and a laser fluence of 2 J/cm$^2$. Square Pt top electrodes 40 nm thick and 50 $\mu$m wide were then deposited using conventional photolithography techniques.

The bottom electrode of the capacitor was connected to ground, and current-voltage (*I-V*) curves were measured. During the forming (set) process, the pristine state (the HRS) is switched into the LRS. The current that flows during this process can result in damage to the capacitors in the circuit. To prevent such damage, we limited the current to a value termed the compliance current, $I_{comp}$. For *I-V* measurements, we used a semiconductor parameter analyzer (Agilent 4155C; Agilent Technologies) and a transistor (2N2369) as the current-limiter. We kept the connection between the bottom electrode and the transistor as short as possible.[8] Using this method, we could set $I_{comp}$ to be as small as 0.7 mA, and the maximum value of $I_{comp}$ used was 20 mA.

We found that $I_{comp}$ was a critical parameter for determining $I_R$. As shown in Fig.

1(a), $I_R$ can be reduced in undoped STO capacitors by decreasing $I_{comp}$. Similar observations have been reported previously, but without physical explanations.[12-14] To explain the relation between $I_{comp}$ and $I_R$, we performed computer simulations using the random circuit breaker (RCB) network model.[3,4] As shown in Fig. 1(b), the RCB network model treats the switching medium as a network composed of circuit breakers with bistable resistance states, which are reversibly switchable using an applied voltage. Details of this model and simulations are described elsewhere.[3,4] This new type of percolation model has been quite successful in explaining numerous properties of unipolar RS, including the nonlinear *I-V* curves observed in the LRS,[4] wide distributions of reset voltages and currents,[3,15] and large 1/*f* noise.[16] Using the RCB network simulations, we could explain why $I_R$ is closely related to $I_{comp}$. Figures 1(c) – (e) show simulation results for network regions where the highly conducting percolating channels were developed during the forming (or set) process. With a smaller value of $I_{comp}$, the volume of the conducting filaments was reduced, and so $I_R$ was also reduced.

There have been some reports of a reduction in the leakage current of STO thin-films by acceptor doping with Co or Mn for high-*k* dielectric applications.[17-19] The leakage current will set the minimum value of $I_{comp}$. Therefore, we proposed that the reduction of the leakage current using acceptor doping could help to decrease $I_{comp}$, which will result in a reduction of $I_R$.

All of the capacitors (i.e., undoped STO, Co:STO, and Mn:STO) showed reliable unipolar RS (Fig. 2(a)), and were highly insulating in the pristine state. The forming process occurs at approximately 6 – 8 V, with $I_{comp}$ = 15 mA, for all the capacitors. With $I_{comp}$ fixed, the set processes occur at 2 – 4 V for most capacitors. This relatively large fluctuation in the set voltage is a well-known characteristic of unipolar RS.[3,15] It should be noted that, with $I_{comp}$ = 15 mA, $I_R$ was quite similar in all the capacitors irrespective of doping.

Figure 2(b) shows *I-V* curves of the capacitors in the pristine state. Note that the leakage current of the STO capacitors can be reduced by about one order of magnitude by Co doping, and doping with Mn results in a further reduction of the leakage current. For comparison, we also fabricated Nb-doped STO (Nb:STO) capacitors using 0.5 % wt. Nb:STO target and the same fabrication conditions. Nb is a donor in STO,[18] and the Nb:STO capacitors showed three orders of magnitude greater conduction than the undoped STO capacitors, as shown in Fig. 2(b). These results indicated that the leakage current in STO capacitors can be varied by controlling the doping, and smaller leakage currents result in larger operating windows in terms of $I_{comp}$, and most notably a lower minimum $I_{comp}$.

To achieve reliable forming, $I_{comp} > 10$ mA was required with the undoped STO capacitors. However, we were able to form conducting filaments reliably in the Co:STO capacitors with $I_{comp} = 2$ mA and in the Mn:STO capacitors with $I_{comp} = 0.4$ mA. The range of $I_{comp}$ required to achieve forming is shown in the right side of Fig. 2(b). Note that the minimum $I_{comp}$ in the Mn:STO capacitors was lower than that of undoped STO capacitors by a factor of about 20. The forming voltages remained in the range 6 – 8 V, and were not significantly affected by the doping.

Figures 3(a) – (c) show *I-V* curves in the LRS obtained with different values of $I_{comp}$ for the undoped STO, Co:STO, and Mn:STO capacitors. For all the capacitors, $I_R$ was reduced by decreasing $I_{comp}$. Despite the difference in doping, the *I-V* curves are quite similar for a given $I_{comp}$. Figure 3(d) shows the relationship between $I_R$ and $I_{comp}$. We can make three important observations. (1) All of the $I_R$ data can be mapped onto a single linear relationship, and $I_R$ is proportional to $I_{comp}$. (2) Irrespective of doping, $I_R$ remains almost constant for a fixed $I_{comp}$. (3) With acceptor doping in the *n*-type STO material system, we can reduce $I_R$ by a factor of at least 20. This enhancement factor is limited by the precision of our measurement system, and so $I_R$ in the Mn:STO capacitors may actually be smaller than the measured values.

It is important to understand how the acceptor doping in STO can allow us to obtain a smaller $I_R$. One possibility is that the electron concentration in the conducting filament will decrease with acceptor doping.[19] If the dielectric constant of the material is assumed not to vary with doping, this will result in a decrease in the current in the percolating channel during the forming or set process, and consequently we will see a reduction in $I_R$. The other possibility is that, when acceptor dopants are introduced and $I_{comp}$ is reduced, the volume of the percolating conduction channels can be decreased, such as in Figs. 1(c) – (e). In our earlier work, we showed that reset voltage should decrease (increase) with reduction of $I_R$ when the percolating channel was multiply (singly) connected.[15] As shown in Figs. 3(a) – (c), the behavior that indicates single percolating channels only appears when $I_{comp} \leq 5$ mA, suggesting a decrease in the percolating channel volume. Both of these physical mechanisms can explain the same observed phenomena, and it is likely that both occur. Further systematic investigations are required to determine how doping can reduce $I_R$ in STO, and whether this doping method will work for other unipolar RS materials.

In summary, we have shown that acceptor doping can significant reduce the reset current in SrTiO$_x$ capacitors. This study suggested that the decrease of compliance current by carrier doping is a viable means of reducing the reset currents in RRAM devices.

This research was supported by the Basic Science Research Program through the National Research Foundation of Korea (NRF) funded by the Ministry of Education, Science, and Technology (Grant No. 2009-0080567). J.S.L. and B.K. were supported by the KOSEF grant funded by the MOST (Grant No. R17-2007-073-01001-0). B.S.K. was supported by the research fund of Hanyang University (HY-2009-N).

FIG. 1. (Color online) (a) *I-V* curves of undoped STO. The red closed and blue open circles show *I-V* curves for $I_{comp}$ of 20 mA and 15 mA, respectively. (b) Schematic diagram of the RCB network model used in the simulations. The red and black bonds represent the circuit breakers in the on and off states, respectively. The right panel shows the switching conditions of the circuit breakers when driven at a bias of $\Delta v$. Simulated results for the RCB network model with different values of $I_{comp}$: (c) 3 a.u. (arbitrary units), (d) 2 a.u., and (e) 1 a.u. The thick red lines show the volume of the conducting filaments, which is reduced when $I_{comp}$ decreases.

FIG. 2. (Color online) (a) Reset and set *I-V* curves, measured from each of the undoped STO (black), Co:STO (red), and Mn:STO (blue) capacitors. (b) *I-V* curves in the pristine state of undoped STO (black), Co:STO (red), Mn:STO (blue), and Nb:STO (green) capacitors. The arrows on the right-hand side indicate the working values of $I_{comp}$ in our experimental setup.

FIG. 3. (Color online) *I-V* for various different values of $I_{comp}$ throughout the operating window, for (a) undoped STO, (b) Co:STO, and (c) Mn:STO capacitors. The numbers indicate the values of $I_{comp}$ in mA. (d) Relationship between $I_R$ and $I_{comp}$. The dashed

lines show the lower limits of $I_R$, and the red arrow at the bottom indicates the lower limits of $I_{comp}$.

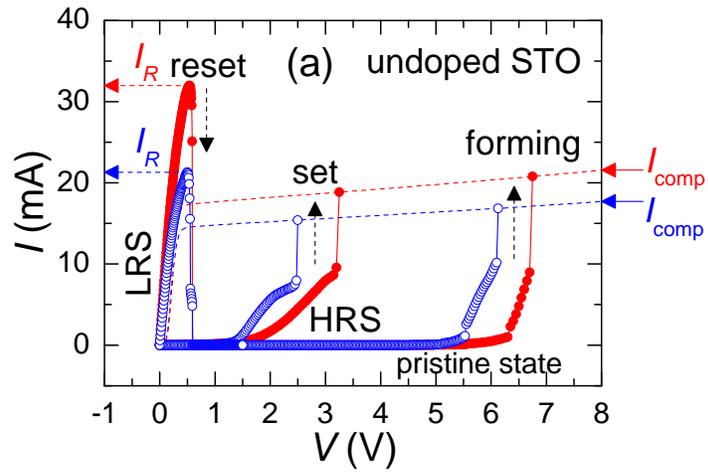
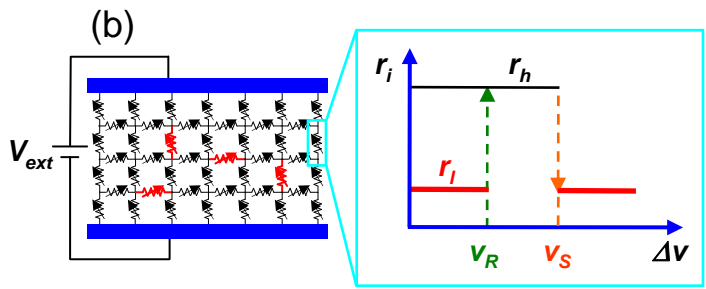
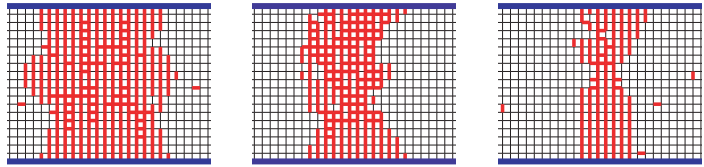

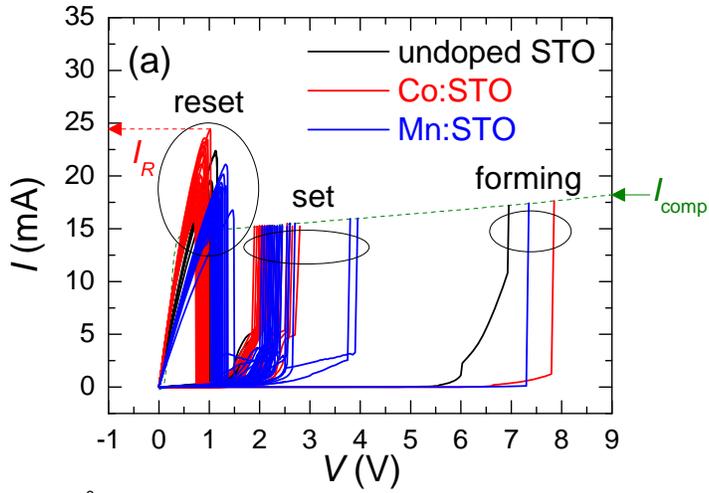
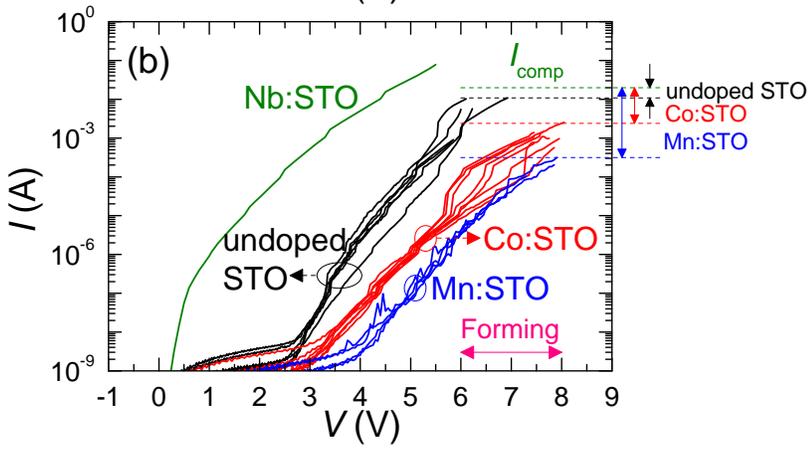

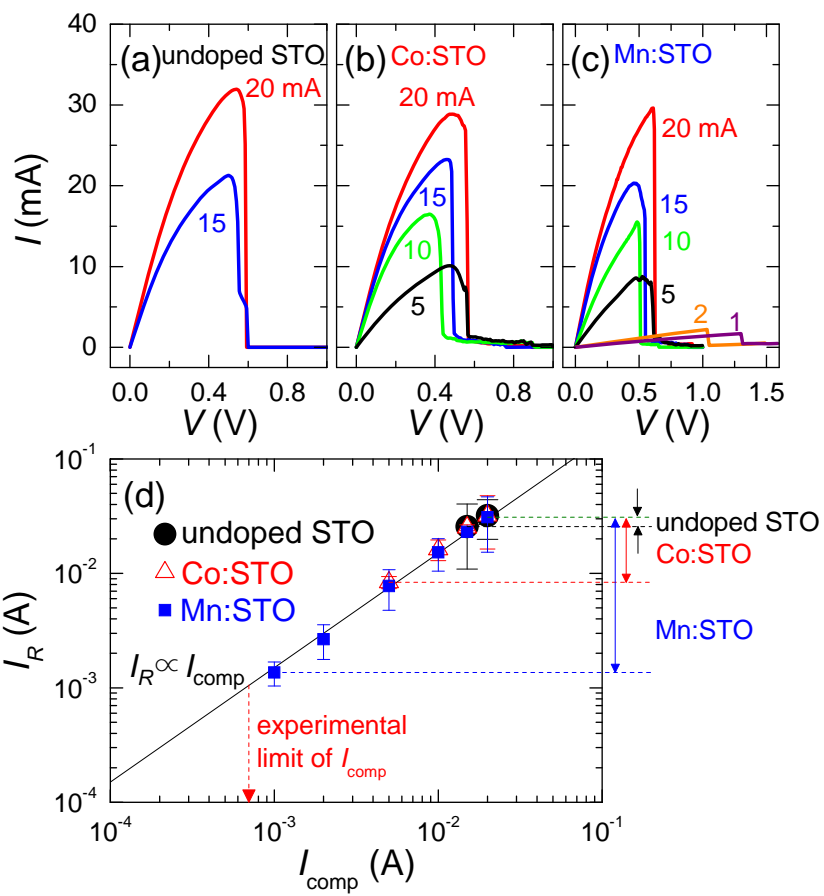